# The Effect of Social Information in the Dictator Game with a Taking Option

Tanya O'Garra[a], Valerio Capraro, Praveen Kujal

Middlesex University London, The Burroughs, London NW4 4BT, UK

[a] Corresponding author: T.Ogarra@mdx.ac.uk

**Abstract**

We experimentally study how redistribution choices are affected by positive and negative information regarding the behaviour of a previous participant in a dictator game with a taking option. We use the strategy method to identify behavioural 'types', and thus distinguish 'conformists' from 'counter-conformists', and unconditional choosers. Unconditional choosers make up the greatest proportion of types (about 80%) while only about 20% of subjects condition their responses to social information. We find that both conformity and counter-conformity are driven by a desire to *be seen* as moral (the 'symbolization' dimension of moral identity). The main difference is that, conformity is also driven by a sensitivity to what others think ('attention to social comparison'). Unconditional giving (about 30% of players) on the other hand is mainly driven by the centrality of moral identity to the self (the 'internalization' dimension of moral identity). Social information thus seems to mainly affect those who care about being *seen* to be moral. The direction of effect however depends on how sensitive one is to what others think.

## 1. Introduction

Understanding whether people's behaviour is influenced by information about what other people have done in a similar context has fascinated generations of economists, psychologists, and sociologists, who have explored this question for decades and from several angles (Asch, 1951; Manski, 1993; Bernheim, 1994; Cialdini & Trost, 1998; Cialdini & Goldstein, 2004; Shang & Croson, 2009).

The overall pattern of results is that people tend to conform to the behaviour of others, arguably because this conveys information about what is normal in a given context and can represent a useful heuristic about how to behave: "if others are doing it, it must be a sensible thing to do" (Cialdini, 1988). Classic social psychology studies have found that people are influenced by what others are doing even in neutral, non-social, settings, such as choosing a consumer product (Venkatesan, 1966) or looking up at the sky (Milgram, Bickman & Berkowitz, 1969). In the domain of social decisions, it has been proposed that people might also conform to avoid social disapproval and gain acceptance from others (Cialdini & Goldsteim, 2004). In social contexts, conformity has been observed both when people receive information about the pro-social behaviour of others and when they receive information about the anti-social behaviour of others. For example, when people are informed about the average voluntary contribution of other people, they tend to conform to this contribution (Shang & Croson, 2009; Bicchieri & Xiao, 2009; Chen, Harper, Konstan and Li, 2010). A similar result holds also when people learn about the anti-social behaviour of others. Social psychology field experiments have found report that people are more likely to litter in a littered setting, compared to a clean one (Finnie, 1973; Reiter & Samuel, 1980; Cialdini, Reno & Kallgren, 1990). Along the same lines, economic experiments using cheating games have found dishonesty to be contagious (Innes and Mitra, 2013; Lauer and Untertrifaller, 2019).

In recent years, however, it has been observed that information about negative behaviour in others can also generate counter-conformity, or reactance, among individuals with strong moral identities. The intuitive logic is that people with strong moral identities seek to avoid feelings of guilt and shame from failing to defend their moral convictions; by counter-conforming they maintain a consistent self-concept (Aramovich, Lytle & Skitka, 2012; Cialdini & Goldstein, 2004). Evidence for counter-conformity has been provided mainly in the context of controversial moral issues: when people with strong moral convictions are informed that they are in minority, they tend to strengthen their convictions (Hornsey, Smith & Begg, 2007; Aramovich, Lytle & Skitka, 2012; Furth-Matzkin & Sunstein, 2017).

We contribute to this literature by studying conformity and counter-conformity using an incentivized economic experiment that proto-typically encapsulates the conflict between a positive, pro-social, action, and a negative, anti-social, action: a dictator game with a taking option (List, 2007; Bardsley, 2008). In our setup, dictators can choose to give money to a recipient, take money from a recipient, or do nothing (neither give nor take). Social information is provided to the dictator prior to making this allocation decision; specifically, they are informed that a previous dictator has either given money to a recipient (positive information), taken money from a recipient (negative information), or neither given nor taken money (neutral information).

Do dictators react to the different pieces of information? If so, do they conform, or do they counter-conform? And what can we say about the personal characteristics - particularly the moral identities - of conformists and counter-conformists? Do counter-conformists have stronger moral identities than other player types? To answer these questions, we collect decisions using the strategy-method (Brandts & Charness, 2011). This allows us to identify the different behavioural strategies that people adopt in response to social information, and hence, to distinguish conformists, counter-conformists and unconditional choosers, based on their individual responses.

We also examine the influence of two key predictors, the "Attention to Social Comparison Information" (AT-SCI) scale (Lennox & Wolfe, 1984) and the "moral identity" scale (Aquino & Reed, 2002). We expect the AT-SCI scale to be an important predictor of conformity in our context, because the scale is meant to measure the extent to which people care about what others think. Lennox and Wolfe (1984) suggest that individuals who have high degrees of sensitivity to what others think tend to avoid negative judgments by others by conforming to what others do. Similarly, we expect the moral identity scale to be another important predictor in our context because both giving behaviour (Krupka & Weber, 2013; Capraro & Rand, 2018) and counter-conformity (Hornsey, Smith & Begg, 2007; Aramovich, Lytle & Skitka, 2012) have been linked to several measures of morality. Therefore, we expect that these correlates might help understand the personality characteristics of people that are more likely to conform (or counter-conform) in response to positive (or negative) information about giving (or taking) behaviour of others.

## 2. Related literature

In the standard 'dictator game', the dictator is endowed with a sum of money and has to decide how much of it, if any, to give to the recipient, who starts the game with nothing. The recipient is passive and only receives the amount that the dictator decides to give. In the dictator game with a taking option, the receiver is also provided with an initial endowment which can be unilaterally taken by the dictator. Dictator games with a taking option were first studied by List (2007) and Bardsley (2008), who found that, in this context, fewer dictators were willing to transfer money to recipients than in the standard game.

Previous work on the effect of social information[1] on dictators' behaviour has mainly focused on standard dictator games, with no taking option. An earlier paper by Cason and Mui (1998) implemented a sequential dictator game in which dictators made an allocation decision before and after receiving social information about another dictator. They found that, whereas irrelevant information led to declines in giving, giving did not change with social information. Bicchieri and Xiao (2009) found that dictator game donations were affected by information regarding the most common behaviour of other dictators (descriptive norm) and by information about what other participants thought it was the most appropriate thing to do (injunctive norm). Krupka & Weber (2009) found that learning about others' behaviour increased donations in the dictator game. Similarly, D'Adda, Capraro and Tavoni (2017) found that informing dictators that about half of the dictators in a previous experiment had donated half of their endowment or more, increased dictator game donations. Servátka (2009) however found that providing information about the behaviour of another dictator did not significantly increase dictator game donations. Zafar (2011) found that subjects changed their donations in the direction of the descriptive norm (what others were doing) in a dictator game with a charitable cause as the recipient.

---

[1] Social information effects are examined in the literature using different names, including "peer effects", "social influence", "neighbourhood effects" and "conformity".

There is also an extensive literature using field experiments to examine the impact of social information on donations to charitable causes. For example, Shang & Croson (2009) found that providing potential donors to a public radio station with information about high contributions by other donors had a positive effect on contributions. Frey and Meier (2004) found that students who were informed that 64% of students had donated to a charitable organization were more likely to donate than students who were informed that 46% of the students had donated. Chen, Harper, Konstan and Li (2010) found that below-median contributors increased their voluntary contributions to an online community after being informed about the median contribution of other users, whereas above median contributors decreased theirs. Goeschl, Kettner, Lohse and Schwerien (2018) found that donations to a pro-environmental cause were affected by the descriptive norm. In sum, the previous literature indicates that people mostly tend to conform to social information regarding the pro-social behaviours of other people.[2]

There is rather less experimental evidence regarding the impact of *negative* social information on behaviour compared to positive social information. The only study to date is an unpublished work by Eugen Dimant (2019), which uses a dictator game with a taking option and a charitable cause as the recipient. In this study, players can revise their original contribution decisions to a charity after learning about the contributions of another player. He finds an asymmetric influence of social information, with taking/anti-social behaviour having a larger (negative) impact on average contributions that giving/pro-social behaviour.

More broadly, classic studies in social psychology have uncovered that people are more likely to litter in a littered setting, compared to a clean one (Finnie, 1973; Reiter & Samuel, 1980; Cialdini, Reno & Kallgren, 1990). In the economic literature, the effect of negative social information on behaviour has been examined using cheating games. Innes and Mitra (2013) found dishonesty to be contagious: when subjects were informed that the rate of dishonesty in a population is high, they were more likely to act dishonestly in a sender-receiver game, compared to when subjects were informed that the rate of dishonesty is low. Similarly, Lauer and Untertrifaller (2019) found people to be more likely to lie after seeing a group member lying.

---

[2]Similar results have been obtained also in behavioural domains other than altruistic giving, such as public goods contributions (Bardsley, 2000; Andreoni & Petrie, 2004; Fischbacher, Gächter & Fehr, 2001; Croson, 2007; Shang & Croson, 2009), effort choices in gift-exchange experiments (Gächter, Nosenzo & Sefton, 2013; Thöni & Gächter, 2015), second- and third-party punishment (Ho & Su, 2009; Fabbri & Carbonara, 2017), trust (Mittone & Ploner, 2011), and, in the field, traffic violation (Chen, Lu and Zhang, 2017).

While all of the above studies regard conformity, there is little work on counter-conformity. Previous work has mainly tested the effect of providing information about a majority opinion that is contrary to participants' held beliefs, on attitudes and/or intentions to speak out against the majority view. Hornsey et al. (2003) found that people with strong moral convictions about a social issue were more likely to express counter-conformist intentions when they learned that they were in a minority, whereas those with weaker moral convictions tended to conform to the majority view. Aramovich, Lytle & Skitka (2012) studied the effect of people's moral convictions in resisting conforming to a majority supporting torture. They found that people with strong moral convictions against torture were more likely to resist conforming. Similarly, Furth-Matzkin & Sunstein (2017) observed counter-conformism in response to several governmental policies, and the extent of counter-conformism was predicted by the extent to which antecedent opinions on the matter were fixed and firm. In the only study to examine whether people exhibit counter-conforming behaviour, Hornsey, Smith and Begg (2007) found that, although participants showed an intention to counter-conform, this resolve disappeared when they were asked to actually act on it.

Related to our work is also the literature testing the effect of moral preferences on dictator game donations. Krupka and Weber (2013) found giving behaviour to be partly driven by preferences for doing what others think is the appropriate thing to do. They also found that not giving was considered to be less inappropriate than taking, and this helped explain framing effects in the dictator game when passing from the give frame to the take frame. Capraro and Rand (2018) and Tappin and Capraro (2018) found dictator game donations to be positively correlated to moral choices in the so-called Trade-Off game, suggesting that dictator game giving is primarily driven by preferences for doing what the dictator thinks it is the morally right thing to do. Capraro and Vanzo (2019) found that moral words in the instructions of the dictator game significantly impacted dictators' level of altruism. For example, dictators were less likely to "steal" from the recipient than to "take", despite the fact that, in their experimental context, "stealing" and "taking" had the same economic consequences.

In summary, evidence from dictator games as well as from field experiments suggest that people are influenced by what other people do. In the context of redistribution behaviours (i.e. giving and taking), people tend to conform to social information about positive (pro-social) behaviour by others; what remains less clear is how people behave in response to negative (anti-social) behaviour in others. The main contribution of this paper is to shed light on how negative social information affects behaviour compared to positive social information, and to identify how moral identity influences how different individuals react to positive and negative social information.

## 3. Experimental Design

We use the dictator game with a taking option, since one of our goals is to examine whether anti-social information generates counter-conformity. To do this requires an action set containing an action (i.e., taking) that is considered to be socially inappropriate, or anti-social. The fact that taking in the dictator game is perceived to be socially inappropriate has been shown by Krupka and Weber (2013). We also test the same hypothesis on our particular sample, and find it confirmed (see Online Appendix 1).

Hence, we conduct a two-player dictator game with a taking option and with social information, as follows. Subjects are randomly assigned to groups of two players, in which one player is randomly assigned the role of dictator (player A) and the other assigned the role of 'recipient' (player B). Player A receives an endowment of $1.50, while player B receives $0.50. Player A is then informed that s/he will have the chance to give money, take money, or do neither with respect to the recipient. The script for player A specifically reads:

> *"In your group, you have been randomly selected to receive $1.50. You are player A. The other participant in your group will receive $0.50. This person is player B.*
>
> *You will now have the chance to give or take some money to/from player B. Specifically, the choices available to you are:*
>
> *Take $0.50 from player B /// Neither give nor take to/from player B /// Give $0.50 to player B"*

Thus, the options available to player A present thee clear choices: 1) an 'anti-social' choice, whereby the player maximises her own payoffs (by taking all of player B's endowment), 2) a 'pro-social' choice, whereby s/he eliminates unequal distributions (by giving $0.50 to player B), and 3) a 'neutral' option, by which she chooses to do nothing.

However, before making this decision, player A is informed she/he has been randomly matched with another participant (player C) who has previously made a decision playing a similar dictator game with another recipient, the game Player D. (The decisions of Players C were collected in a previous experiment[3]). Player A is then asked to indicate her/his contribution decisions *conditional on each possible contribution decision* made by the other player A. Specifically, the text reads:

> *"You will be shown each of the other player's possible decisions. You will then indicate whether you prefer to give to, take from, or neither give nor take from player B in your group in response to each possible decision by the other player."*

The reason why we use the strategy method instead of the direct-response method is to be able to distinguish conformists from counter-conformists and unconditional choosers at the individual level. The order of presentation of the possible choices made by Player C is sequential and randomised across all players A. Each time player A receives new information about Player C's possible choice (i.e. the social information), she/he must select a contribution choice in response to each.

Following completion of the distribution task, all players were asked to explain the reason for their choices using an open-ended explanations format. This was then followed by two psychological scales intended to measure key variables that are expected to influence choices in the game: the "moral identity" scale (Aquino & Reed, 2002) and the "attention to social comparison information" scale (Lennox & Wolfe, 1984).

The moral identity scale, developed by Aquino and Reed (2002), is used to identify the mental representation that an individual may have about her/his moral self (Aquino & Reed, 2002). This construct is developed around key moral 'traits' that have been found to strongly predict moral behaviour (see Aquino & Reed, 2002 for a review of the literature). The reason we add the moral identity scale is the following. Previous work suggests that giving in the dictator game is partly driven by preferences for doing what one thinks others think it is socially appropriate (Krupka &

---

[3] In a prior study we collected data of a baseline game, with no information regarding another participant's behaviour. We use this data as Player C in this experiment.

Weber, 2013) and by preferences for doing what one thinks to be the morally right thing to do (Capraro & Rand, 2018; Tappin & Capraro, 2018; Capraro & Vanzo, 2019). Therefore, we expect that both the "importance of being observed to be moral" and the "importance of moral identity to intrinsic self-image" will affect Player A's choice. Aquino and Reed's (2002) moral identity scale allows us to measure both these two motivations, as it consists of two subscales that distinguish between two dimensions of moral identity: the "*internalization*" dimension (which corresponds to the importance of moral identity to intrinsic self-image), and the *"symbolization"* dimension (which corresponds to the importance of being observed to be moral). We also expect that the extent to which Player A responds to information about the choice of Player C is related to the extent to which Player A cares about being seen as moral. Therefore, we expect that the moral identity scale, and especially its symbolization dimension, is, in our experimental context, related to conformity.

The other scale we use - the AT-SCI scale, developed by Lennox and Wolfe (1984) - is used as a proxy measure of preferences for conformity. This scale specifically measures an individual's sensitivity to social comparison. Lennox and Wolfe (1984) suggest that individuals who have high degrees of sensitivity to what others think (i.e. social comparison) tend to avoid negative judgments by others. The AT-SCI measure has been found to moderate the influence of norms on behavioural intentions (Chiou, 1998; Beardon & Rose, 1990) and judgments (Yoon, La Ferle & Edwards, 2016), such that higher ATSCI individuals are more likely to conform to social norms or information. Therefore, we expect high scores in the AT-SCI scale to be related to conformity in our dictator game with information.

We also include some demographic questions. We focus in particular on gender, since previous analyses suggest that females are more altruistic than males (Engel, 2010; Rand et al., 2016; Brañas-Garza, Capraro & Rascón-Ramírez, 2018), on age, because previous work suggests a positive effect of age on giving (Engel, 2010), and on religiosity, because previous work suggests that religious primes increase giving (Shariff & Norenzayan, 2007; Shariff et al., 2016; but see Gomes & McCullough, 2015).

In the Appendix we report the exact empirical instructions, including the moral identity scale and the AT-SCI scale.

## 4. Results

### *4.1. Participants*

The experiment was conducted on Amazon Mechanical Turk (AMT). AMT experiments are simple and inexpensive, because participants play from their own computers or smartphones by completing an online incentivized survey that takes usually less than ten minutes. This allows experimenters to significantly decrease the stakes at play, without compromising the results. Several works have shown that data collected using AMT are of similar quality than those gathered using the standard laboratory (Paolacci, Chandler & Ipeirotis, 2010; Horton, Rand & Zeckhauser, 2011; Berinski, Huber & Lenz, 2012; Goodman, Cryder & Cheema, 2013; Paolacci & Chandler, 2014). Furthermore, compared to standard experiments, AMT studies use samples that are more heterogeneous than the standard laboratory samples, which are typically made of students (Berinski et al., 2012). On the negative side, some studies have highlighted potential issues with collecting data on AMT, including non-comprehension and, more recently, the presence of AMT workers using Virtual Private Servers (VPS) to participate multiple times in an experiment; critically, these workers provide exceptionally low quality data (Dennis, Goodson & Pearson, 2019). To increase data quality, we recruited participants with an AMT approval rate greater than 95% (AMT keeps track of this information and allows experimenters to filter out participants accordingly), we asked participants comprehension questions to make sure that they understood the crucial parts of the experiment (see below for details about the comprehension questions), and we checked for multiple IP addresses and multiple Turk IDs (in case we find duplicates, we keep only the first observation, as determined by their starting date).

Participants were located in the US. After providing informed consent, they were presented with the experimental instructions, followed by two questions testing comprehension. One question asked which choice by Player A would maximize Player A's payoff; the other question asked which choice by Player A would maximize Player B's payoff. Only participants who correctly answered both questions were allowed to complete the experiment; the other participants were automatically excluded. In doing so, we collected a total of 313 player A's and 311 player B's (females = 48%, mean age = 38 years).

*4.2. Mean transfers in response to social information*

We begin by looking at whether the type of information regarding Player C's behaviour (take, do nothing, or give) impacts Player A's mean transfer. Results in Table 1 suggests the existence of a positive trend such that Player A's mean transfers depend positively on the information received about the behaviour of Player C. Results of a repeated measures ANOVA confirm this intuition, by revealing statistically significant differences in mean transfers in response to different information amounts ($F(2, 624) = 5.35, p < 0.005$). Results of a Friedman test (the non-parametric equivalent to the repeated measures ANOVA) confirm these findings (p<0.001) [4].

**Table 1. Mean Player A transfer by social information (n=313)**

| Information about Player C's transfer | Mean transfer of Players A | Std error |
|---|---|---|
| -$0.50 | $0.00 | 0.024 |
| $0 | $0.02 | 0.023 |
| +$0.50 | $0.04 | 0.024 |

*4.3. Distribution of transfers in response to social information*

Next, we explore whether positive, negative or neutral social information impact the distribution of choices of Players A. Figure 1a shows the distribution of transfer choices (take/neither/give) by social information for the entire dataset, with each panel representing a different social information amount. Figure1b shows the distribution only for those players that actually conditioned their transfer choices in some way to the social information provided. These represent 20% of the full sample - the other 80% of players did not condition their transfer choices to the social information provided. As a consequence, the distributions of choices made by Players A tend to be relatively stable when the information regarding Player C's behaviour varies (Figure 1a). If we restrict the analysis to those Players A who do change their choice as a function of Player C's behaviour, we observe greater variability (Figure1b). Results of a Stuart-Maxwell test for marginal homogeneity confirm that the distributions in Figures 1a and 1b are significantly different from each other (*p<0.001*).[5]

---

[4] A more detailed examination of the underlying regression model corresponding to our ANOVA model shows that only pro-social ('giving') information has a positive effect on transfers compared to anti-social ('taking') information (p=0.001). There is no significant difference between the effect of negative and status quo information on mean transfers (p=0.147). Pairwise comparisons (reported in Online Appendix 2) confirm these findings.

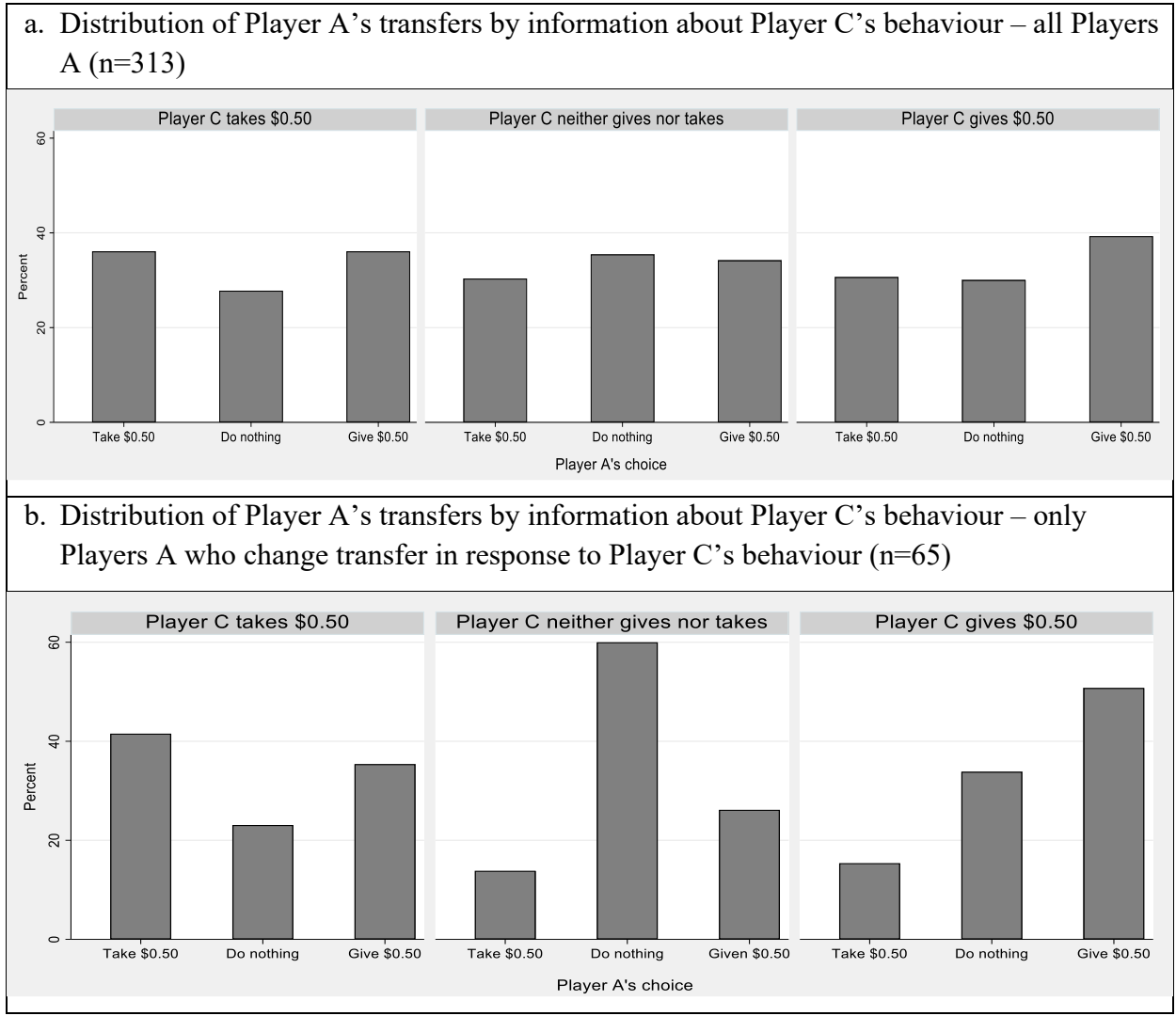

**Figure 1. Distributions of transfers made by Players A by social information.** Panel a. reports the full sample of Players A; panel b. reports only those Players A who change their transfer in response to the social information received about Player C.

---

[5]This result is identical for the full sample and the conditional sample, due to the fact the Stuart-Maxwell test uses only changing frequencies in the calculation.

Having established that the distributions of transfers are statistically different from one another, we now investigate in more detail the impact of each type of information about Player C's behaviour on stated transfers made by Players A. To do this, we focus on the subsample of Players A that actually conditioned their choices on the social information provided (Figure 1b). Table 2 reports results of McNemar pairwise tests.

Starting from the case in which Players A receive neutral information about the behaviour of Player C (central panel of Figure 1b), we note that the majority of these Players A (60% of the conditional subsample) select the same choice, *"neither give nor take",* towards Player B; moreover, the proportion of Players A selecting this choice in response to neutral social information is significantly higher than the proportion of Players A selecting this choice in response to negative (23.1%) and positive (33.9%) social information. This suggests that Players A tend to conform to neutral information about Player C's behaviour.

**Table 2. McNemar tests of proportions of Player A's that condition their responses on social information (n=65).** Significance levels: *: p<0.1, **: p<0.05, ***: p<0.01.

| Comparison | Exact McNemar Test p-value |
|---|---|
| % Players A taking in response to negative information vs % Players A taking in response to neutral information | 0.0005*** |
| % Players A taking in response to negative information vs % Players A taking in response to positive information | 0.0023*** |
| % Players A doing nothing in response to neutral information vs % Players A doing nothing in response to negative information | 0.001*** |
| % Players A doing nothing in response to neutral information vs % Players A doing nothing in response to positive information | 0.0095*** |
| % Players A giving in response to positive information vs % Players A giving in response to negative information | 0.1325 |
| % Players A giving in response to positive information vs % Players A giving in response to neutral information | 0.0113** |

Moving now to Players A who receive positive information about the behaviour of Player C (right panel of Figure 1b), we note that the majority of these players (50.8%) select the same choice, *"give"*, towards Player B; moreover, the proportion of Players A selecting this choice in response to positive information is significantly higher than the proportion of Players A selecting this choice in response to neutral information (26.2%), but not significantly higher than the proportion of Players A selecting the same choice in response to negative information (35.4%). This suggests that, while Players A tend to conform to positive information, responses to negative social information are less conforming.

Indeed, looking now at Players A receiving negative information about the behaviour of Player C (left panel of Figure 1b), we note that this time Players A are more split, as 41.5% decide to take the $0.50 from the corresponding Player B, while a similar proportion (35.4%) decide to give. This shows that some Players A conform to negative information, while a non-negligible proportion of other Players A counter-conform to negative information. Evidence of conformity is confirmed by the fact that the proportion of Players A exhibiting *'taking'* behaviour in response to negative information is significantly higher than the proportion of Players A exhibiting taking behaviour in response to positive (15.4%) and neutral (13.9%) information. On the other hand, the proportion of Players A who *"give"* in response to negative information (35.4%) is numerically but not statistically greater (McNemar test p=0.2632) than the proportion of Players A who *"give"* in response to neutral information (26.2%). This suggests that counter-conformity is relatively rare[6]. We expand on this in the next section.

*4.4. Individual heterogeneity and its determinants*

Summary statistics reported above suggest that there are different responses to social information. In this section, we aim to identify precisely the range of individual strategies that people adopt when responding to social information. To do this, Spearman rank correlation coefficients were computed for each player. This determines the strength and direction of the monotonic relationship between variables, and is the approach used in other studies of player heterogeneity (e.g. Gachter, Gerhards & Nosenzo, 2017; Fischbacher, Gachter & Fehr, 2001). For individuals whose transfer choices did not vary with social information, we could not compute Spearman rank correlations. Hence, mean transfer was used to classify these individuals as 'unconditional givers', 'unconditional takers' and 'unconditional status quo'.

---

[6] If counter-conformity were more prevalent, we would have expected significantly higher levels of giving in response to negative social information compared to neutral or positive social information, as a result of the reactance of players with strong moral identities in response to socially inappropriate behaviour.

We thus define six subject 'types' based on their behavioural strategies, as follows:

- "Conformists": Players A displaying a (weakly) positive correlation between their transfer and the information received regarding the transfer made by Player C;
- "Counter-conformists": Players A displaying a (weakly) negative correlation between their transfer and the information received regarding the transfer made by Player C;
- "Unconditional takers": Players A who take the $0.50 from Player B, regardless of the information received about Player C's choice;
- "Unconditional givers": Players A who give the $0.50 to Player B, regardless of the information received about Player C's choice;
- "Unconditional status quo": Players A who do nothing, regardless of the information received about Player C's choice;
- "Others": Players A who do not belong to any of the previous classes.

Table 3 reports the distribution of types. Results show that about 80% of players did not condition their choices in response to social information about Player C's choices. This leaves only 20.76% of all players that actually conditioned their transfer choices in response to social information. Of these, about 15% consistently conditioned their behaviour to social information, most of whom displayed conformist strategies (10.86%). Only 5.43% displayed counter-conformist strategies, confirming the earlier suggestion that counter-conformity is a rare behavioural strategy. The remaining 4.47% did not show consistent behaviour.

**Table 3. Distribution of subject types**

| Types | Freq. | Percent |
|---|---|---|
| Conformist | 34 | 10.86 |
| Counter-conformist | 17 | 5.43 |
| Unconditional taker | 86 | 27.48 |
| Unconditional status quo | 72 | 23.00 |
| Unconditional giver | 90 | 28.75 |
| Other (u- and inverse-u shapes) | 14 | 4.47 |
| Total | 313 | 100 |

To gain insight on the influence of moral identity and sensitivity to social comparison on the behavioural strategy adopted, we ran multinomial logit regressions, where the dependent variable is player type and the reference type is "unconditional taker". The models include key socio-economic variables (gender, age, and income), religiosity, moral identity (divided into both subscales)[7] and AT-SCI[8] (as an indicator of conformity). In addition, we include the value of the first amount presented to players (recall that Players A's choices were elicited using the strategy method, in which social information regarding the potential choices of Players C was presented sequentially and in random order) to control for possible anchoring effects on the strategy adopted.

Table 3 summarizes the results. Not surprisingly, those classed as 'conformists' tend to score higher on the AT-SCI scale, compared to those who were classed as unconditional takers, suggesting that the likelihood of being a 'conformist' type increases with individual sensitivity to social comparison (Lennox & Wolfe, 1984). With regards to the moral identity subscales, we find that a higher score on the internalization subscale increases the likelihood of being an 'unconditional giver', which is in line with recent empirical evidence showing that giving in the dictator game is primarily driven by moral preferences (Capraro & Rand, 2018; Tappin & Capraro, 2018; Capraro & Vanzo, 2019). Conversely, a higher score on the symbolization subscale increases the likelihood that players select counter-conformist and, to a lesser extent, conformist behavioural strategies. This suggests that both conformism and counter-conformism are primarily driven by a desire to *appear* moral, rather than by an intrinsic moral identity.

In terms of demographic influences, females are more likely than males to adopt unconditional status quo strategies compared to unconditional taking strategies (and, to a lesser extent, unconditional giving strategies). This is partly in line with several meta-analyses showing that females give, on average, more than males in the dictator game (Engel, 2011; Rand et al., 2016; Brañas-Garza et al., 2018). Household income and age have only weak or no statistically significant influences on the likelihood of adopting any single strategy. Also, religiosity has no effect on strategy. This is surprising, because previous research shows that religious primes make people more pro-social (Shariff & Norenzayan, 2007; Shariff et al., 2016; but see Gomes & McCullough, 2015). However, we note that the coefficient of the unconditional givers for religiosity is the only one to be positive, and also the standard error is relatively small; therefore, it is possible that we failed to detect a significant effect of religiosity on unconditional giving, because of insufficient statistical power.

---

[7] We averaged the overall responses to all items and identified within-scale reliabilities using Cronbach's alpha which measures internal consistency, with values closer to 1 being more reliable. Our results show high internal reliability for both the internalization and symbolization scale, with Cronbach's alphas of 0.81 (with mean=6.22 and sd=0.88)) and 0.90 (mean= 4.01, sd=1.45). The overall within scale reliability was 0.86 (overall mean=5.11 and sd=0.98). This indicates that our data are internally consistent.

[8] As with the moral identity scale, we averaged responses over all items (mean 2.40 (sd 0.88)), and assessed within-scale reliability using Cronbach's alpha, which was 0.89. This shows that our data are internally consistent.

Finally, we find that the first amount presented to Player A using the strategy method (the 'anchor'), has a weak influence on the likelihood of adopting a counter-conformist, or 'other' strategy. This confirms findings in O'Garra and Sisco (2019) that anchors may affect the behavioural strategy adopted.

**Table 4. Multinomial Logit Model of Player Types**

|  | Conformist | Counter-conformist | Unconditional status quo | Unconditional giver | Other |
|---|---|---|---|---|---|
| AT-SCI ("conformity" scale) | 0.710** | 0.424 | 0.206 | -0.101 | 0.273 |
|  | (0.290) | (0.357) | (0.202) | (0.196) | (0.366) |
| Moral identity (internalization) | 0.013 | 0.285 | -0.050 | 0.987*** | -0.116 |
|  | (0.258) | (0.402) | (0.192) | (0.255) | (0.375) |
| Moral identity (symbolization) | 0.300* | 0.544** | 0.185 | 0.126 | 0.545** |
|  | (0.170) | (0.231) | (0.135) | (0.130) | (0.249) |
| Female | 0.669 | 0.665 | 0.879** | 0.664* | 1.244* |
|  | (0.450) | (0.586) | (0.348) | (0.350) | (0.638) |
| Age | 0.031* | 0.021 | 0.013 | 0.026* | 0.039 |
|  | (0.018) | (0.025) | (0.015) | (0.015) | (0.025) |
| Household income (/1000) | -0.006 | -0.013 | -0.000 | -0.006 | -0.008 |
|  | (0.006) | (0.009) | (0.004) | (0.005) | (0.009) |
| Religiosity | -0.265 | -0.345 | -0.009 | 0.218 | -0.258 |
|  | (0.268) | (0.349) | (0.207) | (0.205) | (0.379) |
| Anchor: Status Quo | -0.151 | -1.286* | -0.476 | -0.302 | -1.506* |
|  | (0.601) | (0.761) | (0.427) | (0.418) | (0.901) |
| Anchor: Give | 0.389 | -0.895 | -0.385 | -0.388 | -0.604 |
|  | (0.549) | (0.648) | (0.409) | (0.405) | (0.676) |
| Constant | -4.705** | -5.634** | -1.595 | -7.578*** | -4.469* |
|  | (1.683) | (2.561) | (1.166) | (1.671) | (2.321) |
| Chi2 | 95.63*** |  |  |  |  |
| n | 313 |  |  |  |  |

* p<0.1, ** p<0.05, *** p<0.01

## 5. Discussion and Conclusions

We explored how information about the behaviour of a third party affects redistribution behaviour in a dictator game with a taking option. Compared to previous literature, our design proposes three main innovations. First, we did not only look at the effect of positive social information, but also investigated the effect of information about the anti-social behaviour of others. Second, we used the strategy method to examine how individual participants changed their choices as a function of the social information received, which allowed us to identify different behavioural 'types', namely: 'conformists', 'anti-conformists' and unconditional choosers. Third, we explored the determinants of behavioural types using suitable psychological scales, specifically, the "moral identity" scale and the "attention to social comparison information" scale. In doing so, our data provided evidence for a number of novel results.

Whilst the majority of participants (80%) were not affected by social information, about 15% of the participants consistently conditioned their behaviour to the information received; among these, about 10% conformed, and about 5% counter-conformed to the behaviour of a third party. The remaining 5% did not show consistent behaviour. The investigation of heterogeneous individual strategies also revealed some interesting patterns: individuals classed as 'unconditional givers' were more likely to have strongly internalized moral identities (as per the 'internalization' dimension of the moral identity scale). On the other hand, individuals classed as 'conformists' as well as those classed as 'counter-conformists', were mostly influenced by a desire to be *seen* as moral (as measured using the 'symbolization' dimension of the moral identity scale). However, conformists also showed concern for what others think (as measured by the 'attention to social comparison' scale). Thus, our results suggest that social information mainly affects those who care about being *seen* to be moral, at least in what regards redistribution choices. The direction of effect however depends on how sensitive one is to what others think.

These results confirm and go beyond previous literature in several dimensions. In the Related Literature section, we reviewed studies finding that participants who receive positive social information tend to conform. We replicated this finding, and also showed that, when participants receive negative social information (i.e. information about a third-party taking money from recipients in a dictator game), some of them conformed to this negative behaviour, but others counter-conformed. The finding that some participants adopt counter-conformist behaviour in response to negative social information, is, to the best of our knowledge, new, although it relates to the recent line of evidence that people with strong moral convictions exhibit counter-conformist attitudes and intentions in response to information that goes against their prior moral and political beliefs (Hornsey, Smith & Begg, 2007; Aramovich, Lytle & Skitka, 2012; Furth-Matzkin & Sunstein, 2017).

To the best of our knowledge, our study is the first to explore the personality determinants of conformity and counter-conformity, at least in the domain of giving and taking behaviour. We found that, in our context, both conformity and counter-conformity were driven by a desire to be seen as moral. This result can be useful in terms of applications. Future work could test whether moral salience can increase people's conformity to positive behaviours and people's reactance to negative behaviours. However, we also found that conformity was also driven by a sensitivity to what others think. Future work could test whether individuals who are sensitive to social comparison respond equally to social information provided by in-group members and social information provided by out-group members. If explicitly faced with information about the behaviour of out-group members, might these individuals change their behavioural strategy from conformist to counter-conformist? This research can contribute to the emerging stream of literature using moral suasion to encourage pro-social behaviours (Brañas-Garza, 2007; Ferraro & Miranda, 2013; Dal Bó & Dal Bó, 2014; Bonan, Cattaneo, d'Adda & Tavoni, 2019; Bott, Cappelen, Sørensen & Tungodden, in press; Capraro et al., in press).

Finally, our results also confirm previous work showing that giving is partly driven by moral preferences for doing the right thing (Capraro & Rand, 2018; Tappin & Capraro, 2018; Capraro & Vanzo, 2019). We add to this line of research by highlighting that moral preferences may not just increase giving, but indeed may do so in contexts where other people are observed to do quite the opposite.

**Acknowledgements**

We acknowledge financial support for this study from the Research Facilitation Funding scheme and Economics Department at Middlesex University. We also thank Alex Bueno for programming the study, and Joe Miele for distributing the experiment on MTurk. An online appendix, inclusive of experimental instructions, is available.

**Online Appendix**

## 1. Perceived Morality of Different Redistribution Choices

To identify perceptions of the morality of the various possible transfer choices (give, take, neither), we asked Players B to indicate on a scale of 1 to 7 (where 1=extremely moral/bad, and 7=extremely moral/good) how moral they considered each of these choices. This question was asked *prior* to informing them that they had been assigned the role of player B, to avoid bias in terms of how they responded to the questions. We did not ask Players A to avoid influencing their choices. Results show that 'taking' was rated as immoral (mean rating of 2.23), whereas 'giving' was rated as moral (mean rating of 6.22 on the scale). 'Neither giving nor taking' was rated as 4.14, thus perceived as marginally more moral than immoral (assuming a rating of "4" indicates 'neither moral nor immoral'). Paired pairwise t-tests comparing the mean perceived morality of each possible decision indicate that these ratings are all significantly different for each other (all p-values<0.001).

## 2. Testing Difference between Mean Transfers

**Table 5. Pairwise Paired Two-Tailed T-tests and Wilcoxon Signed-rank tests (n=313)**

| Tests of difference between mean transfers by social information | Paired t-test (p-value) | Sign rank test (p-value) |
|---|---|---|
| H$_0$: mean transfer in response to 'take' = mean transfer in response to 'give' | 0.0049*** | 0.0145** |
| H$_0$: mean transfer in response to 'take' = mean transfer in response to 'neither' | 0.1089 | 0.1539 |
| H$_0$: mean transfer in response to 'give' = mean transfer in response to 'neither' | 0.0507* | 0.0354** |

**EXPERIMENTAL INSTRUCTIONS**

Screen 1

--ALL PLAYERS--

**Please read the following information carefully:**

You are invited to take part in a study that investigates choices made by individuals in group settings. Participants will be asked to respond to some questions and make some choices which may affect other participants in this survey. No choices will be traced back to any one individual.

**The following list of items summarises all important things you should know before proceeding with this study.**

1. You must be over 18 to participate.
2. Your participation is voluntary, and you may withdraw from the study at any time.
3. Participation in the tasks will not incur any financial expense by you.
4. Some of the tasks may involve monetary reward. Any earnings will be relatively small and related to the decisions you and other participants make.
5. If you agree to participate in the study, you are expected to fulfil the obligations related to the study. That is, respond to the tasks assigned to you for the duration of the study.
6. There are no known physical risks involved in this procedure and the tasks do not require any special physical or psychological attitudes or any specific knowledge of any kind.
7. You will not be knowingly deceived in any form.
8. During this study we may ask you for some personal information. For instance, your gender, level of education, personal income level, etc.
9. **CONFIDENTIALITY:** The information you provide will be treated in full confidence and will be legally protected. It will never be associated with you personally in any form. No person-identifiable information will be reported in any published or unpublished work. Non-person identifiable data may be made publicly available. All electronic files will be saved but treated in accordance with the Data Protection Act (1998).

**Consent:**
If you agree to the terms of this study and wish to continue with this study, then please proceed to the next page.

Screen2

Thank you for participating in this study.

Important: Please do not use the 'Back' and 'Forward' buttons in your browser.

You will earn a $0.75 participation fee for completing the study. You will also have the chance of earning more money. The extra amount earned will depend on the choices that you and other participants make during the survey.

When you are ready, please move to the next page to start the study.

Screen 3 Explanation Set-up (All players)

For the purpose of this study, you have been randomly assigned to a group of 2 Mechanical Turk participants, including yourself.

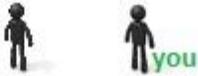

Two bonus payments of $1.50 and $0.50 will be randomly allocated to each ONE of you.

The participant who receives $1.50 will be known as player A.

The participant who receives $0.50 will be known as player B.

If you are randomly selected to receive $1.50, you will have the opportunity to either give some of your own bonus to player B, or take some of the bonus received by player B.

The other person in your group is REAL - there is no deception in this study. This is the only interaction that you will have with the other participant. He or she will not be able to influence your bonus in later parts of the experiment

The following questions are designed to test whether you have read and understood the above scenario.

*You MUST answer the next two questions correctly to continue with the HIT and receive your participant fee and bonus!*

Screen 4 (Comprehension Test 1)

Q1. If you have been randomly selected to be player A, how much should you give or take to/from player B, so that you and player B earn the same amount?

| Amount | Check one only |
|---|---|
| Take $0.50 | |
| Neither give nor take | |
| Give $0.50 | |

GO TO SCREEN 5

Screen 5 (Comprehension Test 1)

Q2. If you have been randomly selected to be player A, how much should you give or take to/from player B to maximize your earnings?

| Amount | Check one only |
|---|---|
| Take $0.50 | |
| Neither give nor take | |
| Give $0.50 | |

IF CORRECTLY ANSWERED Q1 & Q2: GO TO SCREEN 7

IF NOT: END OF PARTICIPATION

Screen 7 Selection (Text)

The computer will randomly select one participant in your group to receive the $1.50 bonus (player A), and one participant to receive $0.50 each (player B).

Please note: you MUST complete the entire study to receive your participation payments and any earned bonus payments.

Please move to the next page.

Screen 8: Role assignment (player A); explanation of decisions faced

In your group, **you have been randomly selected to receive $1.50**. You are player A.

The other participant in your group will receive $0.50. This person is player B.

You will now have the chance to give or take some money to/from player B. Specifically, the choices available to you are:

| Take $0.50 from player B | Neither give nor take to/from player B | Give $0.50 to player B |
|---|---|---|

However, before you make this decision, you will be randomly matched with another player A who has already decided whether to give to, or take from, player B in their group.
You will be shown each of the possible decisions that the other player A can make.

You will then indicate whether you prefer to give to, take from, or neither give nor take from player B in your group conditional on each possible decision by the other player A.

Please note that all your decisions are potentially binding and cannot be changed.

Please move to the next page.

GO TO SCREEN 10

Screen 10 Transfer decision (Player A)

Now, please indicate how much money you wish to give to or take from player B in your group conditional on the other player A's possible choices.

*Please note: all transfer decisions are confidential and will not be traced back to you.*

-- RANDOMISE OPTIONS --

If **the other player A chose to GIVE $0.50** to player B in their group, then I choose to:

(select one):

| Take $0.50 from player B in my group | Neither give nor take to/from player B in my group | Give $0.50 to player B in my group |
|---|---|---|
|  |  |  |

If **the other player A chose to NEITHER GIVE NOR TAKE** to/from player B in their group, then I choose to:

(select one):

| Take $0.50 from player B in my group | Neither give nor take to/from player B in my group | Give $0.50 to player B in my group |
|---|---|---|
|  |  |  |

If **the other player A chose to TAKE $0.50** from player B in their group then I choose to:

(select one):

| Take $0.50 from player B in my group | Neither give nor take to/from player B in my group | Give $0.50 to player B in my group |
|---|---|---|
|  |  |  |

Screen 11

You have now made your decisions conditional on the other player A's choices. You will find out the other player A's choice - and hence which of your decisions applies - when you receive your payment after this experiment.

Screen 12 Reasons for transfer decision

**Can you explain in a few sentences how you made your decisions?**

|  |
|---|
|  |

SCREEN 13: MORALITY OF BEHAVIOUR QUESTION

**-- ALL PLAYER B's ONLY --**

Before you find out which role you have been randomly assigned to, please indicate on a scale of 1 to 7, where 1=extremely immoral/bad and 7=extremely moral/good, how *moral* you consider the following behaviours:

|  | extremely immoral/bad 1 | 2 | 3 | neither moral nor immoral 4 | 5 | 6 | extremely moral/good 7 |
|---|---|---|---|---|---|---|---|
| Taking $0.50 from player B |  |  |  |  |  |  |  |
| Neither give nor take to/from player B |  |  |  |  |  |  |  |
| Giving $0.50 to player B |  |  |  |  |  |  |  |

Screen 14: Role assignment; player B text

-- ALL PLAYER **B**'s --

In your group, **you have been randomly selected to receive $0.50**. You are player B.

The other participant in your group will receive $1.50. This person is player A.

Player A will now have the chance to give or take some money to/from you.

You will learn how much they chose to give or take when you receive your payment after completing the study.

-- ALL PLAYERS FROM SCREEN 15 ONWARDS–

Screen 15: MORAL IDENTITY

Listed below are some characteristics that might describe a person:

*caring, compassionate, fair, friendly, generous, helpful, hardworking, honest, kind.*

The person with these characteristics could be you or it could be someone else.

For a moment, visualize in your mind the kind of person who has these characteristics. Imagine how that person would think, feel, and act. When you have a clear image of what this person would be like, answer the following questions.

*For each question below: 7-point Likert scale ranging from 1 (strongly disagree) to 7 (strongly agree)*

| |
|---|
| 1. It would make me feel good to be a person who has these characteristics. |
| 2. Being someone who has these characteristics is an important part of who I am. |
| 3. I often wear clothes that identify me as having these characteristics. |
| 4. I would be ashamed to be a person who had these characteristics. |
| 5. The types of things I do in my spare time (e.g., hobbies) clearly identify me as having these characteristics |
| 6. The kinds of books and magazines that I read identify me as having these characteristics. |
| 7. Having these characteristics is not really important to me. |
| 8. The fact that I have these characteristics is communicated to others by my membership in certain organizations. |
| 9. I am actively involved in activities that communicate to others that I have these characteristics. |
| 10. I strongly desire to have these characteristics. |

Screen 17: CONFORMITY (ASCI scale)

Now, please indicate your level of agreement with the following statements:

*(For each question below, 5-point Likert scale: 0 = completely disagree to 5=completely agree).*

| 1 | It is my feeling that if everyone else in a group is behaving in a certain manner, this must be the proper way to behave. |
|---|---|
| 2 | I actively avoid wearing clothes that are not in style. |
| 3 | At parties I usually try to behave in a manner that makes me fit in. |
| 4 | When I am uncertain how to act in a social situation, I look to the behavior of others for cues. |
| 5 | I try to pay attention to the reactions of others to my behavior in order to avoid being out of place. |
| 6 | I find that I tend to pick up slang expressions from others and use them as part of my own vocabulary. |
| 7 | I tend to pay attention to what others are wearing. |
| 8 | The slightest look of disapproval in the eyes of a person with whom I am interacting is enough to make me change my approach. |
| 9 | It's important to me to fit in to the group I'm with. |
| 10 | My behavior often depends on how I feel others wish me to behave. |
| 11 | If I am the least bit uncertain as to how to act in a social situation, I look to the behavior of others for cues. |
| 12 | I usually keep up with clothing style changes by watching what others wear. |
| 13 | When in a social situation, I tend not to follow the crowd, but instead behave in a manner that suits my particular mood at the time |

Screen 18: Gender

Thanks! A few more questions about yourself:

Are you..? (*Check one only*)
[ ] female
[ ] male
[ ] other

Screen 19: Age

In what year were you born?
…………

Screen 20: Education
What is the highest level of education you have received?
*Check one only*
[ ] Less than high school
[ ] High school/ GED
[ ] 2-year college degree
[ ] 4-year college degree
[ ] Master's degree
[ ] Doctoral degree
[ ] Professional degree (JD, MD)

Screen 21: Household income
What is your combined annual household income, before tax? *Please remember that all answers are confidential. Income is a very useful measure for research purposes.*
*Check one only*
[ ] Less than $30,000
[ ] $30,000-$39,999
[ ] $40,000-$49,999
[ ] $50,000-$59,999
[ ] $60,000-$69,999
[ ] $70,000-$79,999
[ ] $80,000-$89,999
[ ] $90,000-$99,999
[ ] $100,000-$124,999
[ ] $125,000-$149,999
[ ] $150,000 or more

Screen 22: political tendency
Do you consider yourself to be a Democrat, Republican, Independent or Other?
*Check one only*
[ ] Democrat
[ ] Republican
[ ] Independent
[ ] Other

Screen 23: Religiosity

How strongly do you believe in the existence of a God or Gods?

1 = not at all
2 = somewhat
5 = very much

38